\documentclass[amsmath,amssymb,twocolumn,pre,floatfix]{revtex4}

\def\D{{\rm d}}

\usepackage{color}
\usepackage{graphicx}
\usepackage{dcolumn}
\usepackage{times}
\usepackage{soul}

\usepackage{afterpage}

\begin{document}

\title{Active textiles with Janus fibres}
\author{A.~P. Zakharov, L.~M. Pismen}
\email{pismen@technion.ac.il}
\affiliation{Department of Chemical Engineering, 
Technion -- Israel Institute of Technology, Haifa 32000, Israel}

\begin{abstract}
We describe reshaping of active textiles actuated by bending of \emph{Janus fibres} comprising both active and passive components. A great variety of shapes, determined by minimising the overall energy of the fabric, can be produced by varying bending directions determined by the orientation of  Janus fibres. Under certain conditions, alternative equilibrium states, one absolutely stable and the other metastable coexist, and their relative energy may flip its sign as system parameters, such as the extension upon actuation, change. A snap-through reshaping in a  specially structured textile reproduces the Venus flytrap effect.   

\end{abstract} 
 \maketitle

\section{Introduction} 

One of the novel ideas in the rapidly developing field of biomimetic materials is the use of textiles \cite{textile} or printed grid patterns \cite{print4d} for creating variable forms. Extension or contraction of \emph{active} fibres made of a longitudinally polarised nematic elastomer due to a change of the nematic order parameter causes textiles or printed grids to acquire a wide variety of shapes. Alternatively, elongation may be caused by local swelling of an elastic hydrogel  \cite{Balazs} and shortening by expelling the solvent. The two cases differ only by a change of the fibre radius $r$, which grows in the same proportion as the length in the hydrogel but shrinks to preserve the volume in nematic elastomers. We will further concentrate on the nematically actuated textiles but only slight changes have to be made to adjust to the swelling case. Other ways of activation, \emph{e.g.} electrostriction may be employed. 

\begin{figure}[b]
	\begin{tabular}{cccc}
		(a)&(b)\\
			 \includegraphics[width=0.245\textwidth]{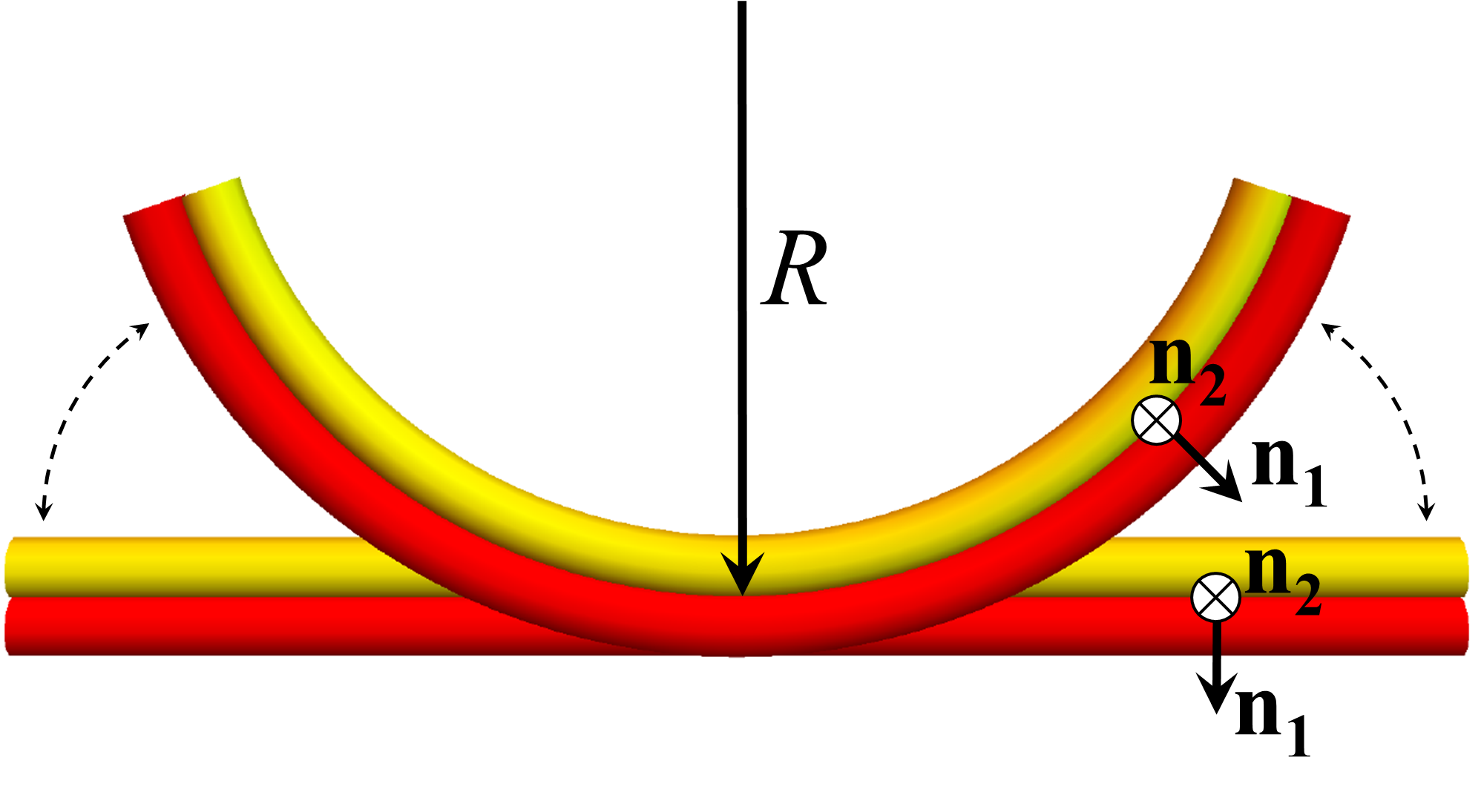}
			& \includegraphics[width=0.22\textwidth]{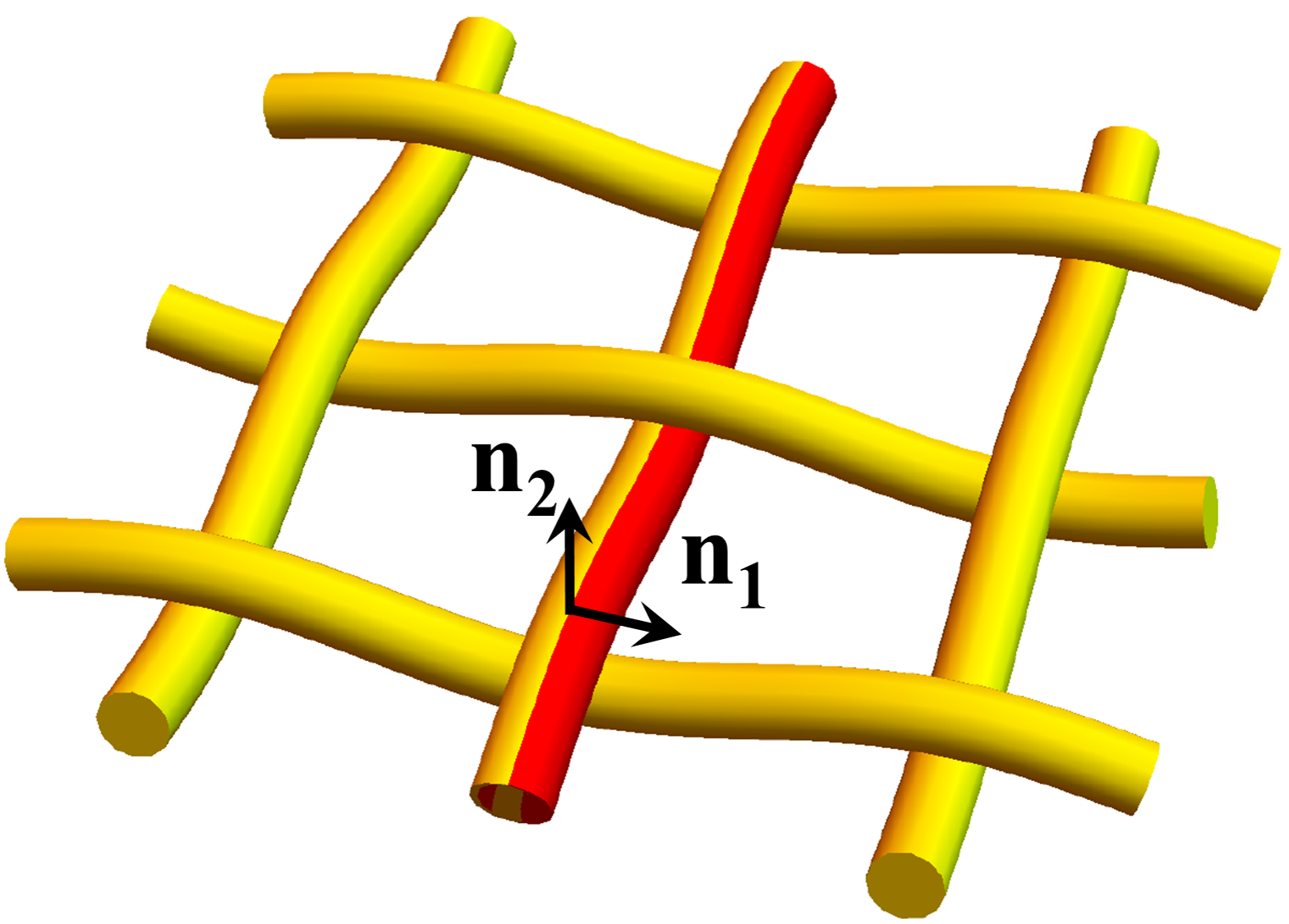}\\
	\end{tabular}
	\caption{(a)A Janus-fibre made of two linked filaments before and after actuation.  (b) A piece of textile made as a woven regular structure with passive fibres (light, or yellow) and an embroidered active fibre (mixed shades or colors) bending in the textile plane. }
	\label{f:fibre}
\end{figure}

The advantage of textiles or grids, as compared to deformable thin sheets, is better versatility due to the absence of continuity constraints that lead to ubiquitous buckling patterns under variable local strain. Particularly interesting effects may be generated by fibres acquiring spontaneous bending with a change of nematic order. Those are \emph{Janus} filaments, which can be fabricated using two connected extruders with simultaneous melt spinning to generate a fibre containing two different materials, say, a nematic and isotropic elastomers \cite{IonovJanus}. If the length of the nematic component changes, spontaneous curvature arises in the direction $\textbf{n}_1$ determined by its orientation, as shown in Fig.~\ref{f:fibre}a. The direction of bending depends on the orientation of the dividing plane and varies in a twisted fibre. The same effect can be achieved by gluing side-by-side two elastomer filaments, one nematic and one isotropic. Twisted yarns of this kind featured recently as a nematoelastic crawler \cite{azlp} and as a flexible Stokesian swimmer \cite{swim}. 

The direction of curvature can be chosen at will, which gives an advantage over continuous bilayer or twisted-nematic films where the curvature can be oriented only normally to the surface. In-plane deformations may appear due to spatial inhomogeneities, caused, for example, by phase separation upon swelling \cite{SoftMatter2017} but cause intensive buckling due to a higher cost of extension compared to bending. In textiles, unlike continuous films, bending is governed directly by orienting Janus fibres, and buckling is avoided since no area or volume conservation conditions are imposed on a woven textile. 

The most interesting feature of active textiles considered below is the possibility of a fast transition between different shapes. Due to the absence of area or volume conservation, such transitions are achieved much easier than in continuous sheets, and should occur faster due to a low aerodynamic resistance to reshaping of a fibre network. This allows us, in particular, to reproduce a ``Venus flytrap" effect \cite{MahadevanFlytrap}.

In this paper, we consider reshaping effects generated in textiles incorporating active Janus fibres. We will explore just a sample of a great variety of arrangements: combining active and passive filaments spun in different directions; weaving and stitching in different ways; framing or embroidering a passive textile by active fibres; actuating active fibres independently or simultaneously, coiling or twisting them in different ways, changing extension coefficients and Young moduli or radii that affect flexural rigidity. More variety can be achieved by varying the structure and shape of the mesh, which can be designed as a flat strip, a curved surface or even extended to a multi-layered structure or a three-dimensional network. 

Out of this variety, we restrict to actuation of flat square pieces of textile loosely woven as shown in Fig.~\ref{f:fibre}b, which allows for sliding of fibres at intersection nodes. Following the description in Sect.~\ref{S2} of the geometry and elastic energy of filaments incorporated in the computation algorithm, we consider in Sect.~\ref{S3} effects of in-plane actuation specific for textiles, paying particular attention to transitions between alternative structures. Off-plane actuation, including construction of a Venus flytrap is considered in Sect.~\ref{S4}.

\section{Geometry and energy of active textile} \label{S2}

We consider in detail textiles actuated by Janus fibres having prior to actuation a circular cross-section with the radius $r_a$. When the active component of the Janus fibre elongates locally by a factor $\lambda = 1+\epsilon$, it bends upon actuation with a curvature radius $R$ along the vector $\textbf{n}_1$ normal to the dividing plane. We will presently see that the ratio $r_a/R=O(\epsilon)$ at $\epsilon \ll1$. The common thin string approximation is applicable at $r_a \ll R$, and therefore we further restrict to small extensions.

Let the active component occupy a half-circle with the angular coordinates $|\theta|<\pi/2$. Upon actuation, the length of the fibre extends by the factor $1+ \epsilon/2$ and the total cross-sectional area of the fibre decreases by the same factor. as the area of the active part shrinks by the factor $\lambda$. The sector occupied by the active elastomers reduces thereby to $|\theta|<\pi/2(1-\epsilon)$, which introduces negligible highs-order corrections at $\epsilon \ll1$. In a naturally bent fibre, the passive sector will be on the inner, and the active one on the outer side of the arc, and both parts will be under compression relative to their zero energy state. Deformation along the fibre axis at a point with the coordinates $r,\, \theta$ is $u =(r/R) \cos\theta$. In the active part, the strain relative to the intrinsic extension is $u-\epsilon$. The total energy per cross-sectional area $A=\pi r_a^2$ obtained by integrating is $E A/2 (I_\mathrm{in}+I_\mathrm{out})$, where $E$ is the Young modulus and $I_\mathrm{in},\,I_\mathrm{out}$ are partial momenta for the active and passive sectors computed as area integrals of, respectively, $u^2$ and $(u-\epsilon)^2$. A short computation yields, up to $O(\epsilon^3)$ corrections,
\begin{align}
I_\mathrm{in} &=\frac{2}{\pi r_a^2}\int_0^{r_a} r \,\D r 
\int_0^{\pi/2} \left(\frac rR \cos \theta -\epsilon\right)^2 \D \theta  \notag \\
&= \frac 18  \left(\frac{r_a}{R}\right)^2 - \frac{4\epsilon}{3\pi}\frac{r_a}{R} +  \frac{\epsilon^2}{2} ,
  \label{Iin} \\
I_\mathrm{out} &=\frac{4}{\pi r_a^2}\int_0^{r_a} \frac {r^3}{R^2} \,\D r  
\int^\pi_{\pi/2}\cos^2 \theta \,\D \theta
= \frac 18  \left(\frac{r_a}{R}\right)^2, \label{Iout} \\
I_a & = I_\mathrm{in}+I_\mathrm{out} = \frac 14  \left(\frac{r_a}{R}\right)^2
 -\frac {4\,\epsilon}{3 \pi} \frac { r_a}{R}+\frac {\epsilon^2}{2} ,
 \label{Ipi}
\end{align}
The natural curvature radius of the fibre upon actuation $R_0$ is determined by the condition of minimum overall strain energy in the cross-section, yielding $R_0/r_a =8/(3\pi \epsilon) \approx 0.849/\epsilon$. The residual energy at the optimal bending is of order $O(\epsilon^2)$.  The momentum of passive fibres is twice Eq.~\eqref{Iout} with $r_a$ replaced by $r_p$, which recovers the standard expression $I_p= \frac {1}{4}(r_p/R)^2 $. 

To keep the integrity of the fabric, it should be stitched at edges. We assume that stitching does not restrict rotation of internal fibres at the boundary nodes. This weaving and stitching pattern excludes torques at intersection nodes, which prevents filaments from twisting upon actuation. The woven structure of the textile in Fig.~\ref{f:fibre}b imposes a ``microcurvature" component along the normal to the envelope surface of the fabric, that depends on the distance $\ell_{i,j}$ between intersections and prevents convergence of neighbouring nodes to distances comparable to the diameter of the fibres. The microcurvature $\kappa_{i,j}$ at a node $(i,j)$ of the $i$th filament is approximated in the limit $r \ll \ell_{i,j}$ by $\kappa_{i,j} \approx 4r/(\ell_{i,j+1}+\ell_{i,j-1})$.  In actual computations, this additional curvature is taken into account by shifting the locations $x_{i,j}$ of intersecting filaments by $r$ along the normal to the envelope of the fabric and using the usual approximant $\kappa_{i,j} =2 |\alpha_{i,j}| /(\ell_{i,j+1}+\ell_{i,j-1})$, where $\alpha_{i,j}$ is the angle between the two adjacent segments. The approximation is valid as long as the local curvature radius is large compared to the distance between nodes. The intersecting filaments should remain in contact under these conditions, since their separation will increase the microcurvature, which is minimal (though nonzero) when the network is regular and increases with growing inhomogeneities. 

The elastic energy of a deformed thin filament is a combination of pure extension (stretching) and bending, with torsion excluded due to the absence of torques. We define the elastic energy of fibres per unit length $\mathcal{F}_e$ as 
\begin{equation}
\mathcal{F}^e =  \frac12 E A \left( \mathcal{F}^s +\mathcal{F}^c \right), \quad
\mathcal{F}^s = u_z^2,  \quad  \mathcal{F}^c= I(R),
\label{Fe}
\end{equation}
Here $u_z$ is the axial strain of the filament, and the momenta $I(R)$ are $I_a$ as defined by Eq.~\eqref{Ipi} for active, and $I_p= \frac {1}{4}( r_p/R)^2 $ for passive filaments. Since the extensions have to be small, slight variations of the cross-sectional area can be neglected. The stretching energy $\mathcal{F}^s $ is computed as a sum over filaments. Since gliding on intersections is allowed, the axial strain $u_z$ of each fibre (which is suppressed at $r \ll R$) is defined as the deviation from the original arc length between the stitched ends, and therefore it depends only on their positions $\mathbf{x}_{i}^\pm$. The change of the total length $L_i$ due to varying microcurvatures at the nodes can be neglected. 

The equilibrium configuration minimising the total energy of the filament network is attained following the pseudo-time evolution equations for the positions of nodes $\mathbf{x}_{ij} $:
 \begin{equation}
\frac{\D \mathbf{x}_{i,j}}{\D t}= - \frac{\partial}{\partial \mathbf{x}_{i,j}} \left( 
 \sum_\mathrm{nodes}\mathcal{F}^{c}_{i,j}
 + \sum_\mathrm{fibres}\mathcal{F}^{s}_{i} \right) .
 \label{evol}
\end{equation}
\section{In-plane actuation \label{S3}}

Janus fibres can be oriented at any angle $\gamma$ to the original plane of the fabric. The simplest case is in-plane orientation with $\gamma$ equal to 0 or $\pi$. Even then, planar shape can become either impossible or unstable after actuation. Such kind of actuation is particularly interesting, since it is specific to textiles, and cannot be realised in a continuous sheet. 

\subsection{A model example}

%
\begin{figure}[b]
\begin{tabular}{cccc}
		(a)&(b)&(c)\\
	  \includegraphics[width=0.15\textwidth]{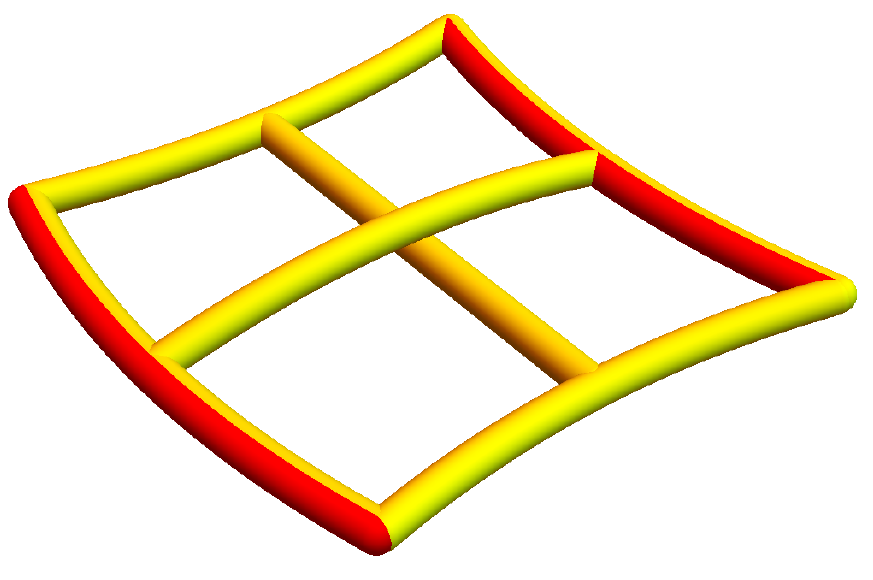}
	& \includegraphics[width=0.15\textwidth]{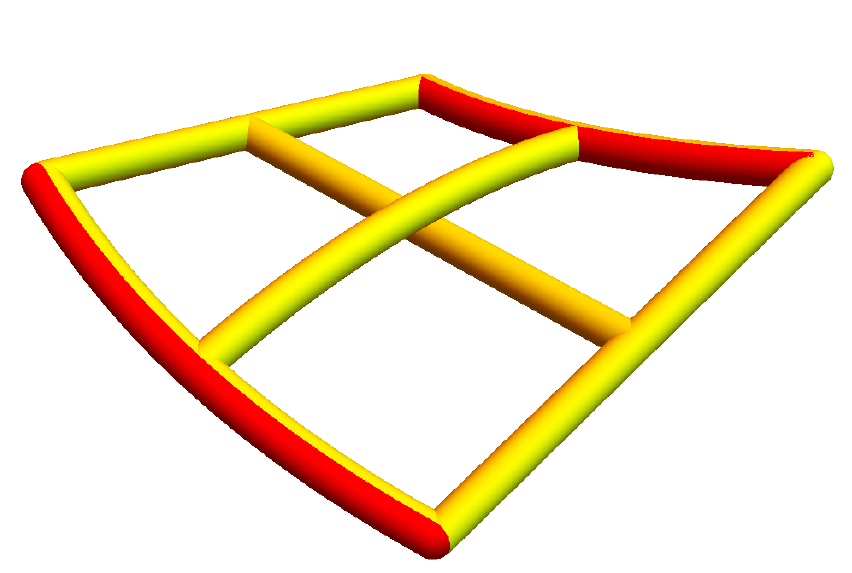}
	& \includegraphics[width=0.15\textwidth]{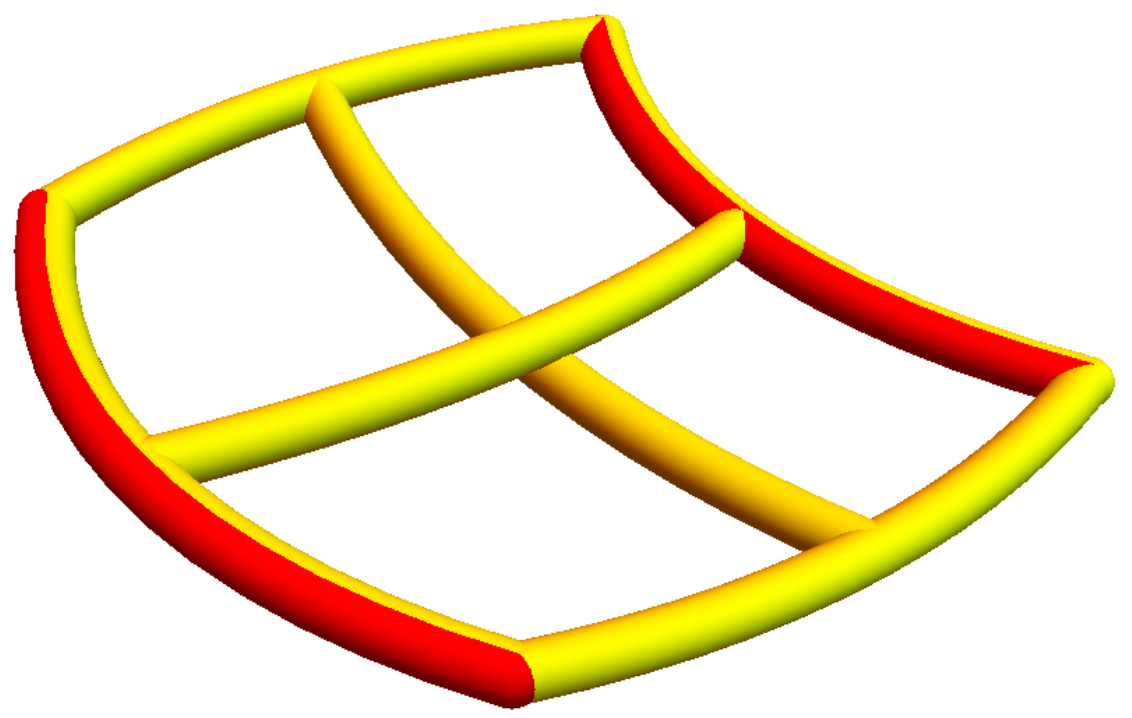}
\end{tabular}\\ \vspace{3mm}
	(d)\\ 
\includegraphics[width=0.45\textwidth]{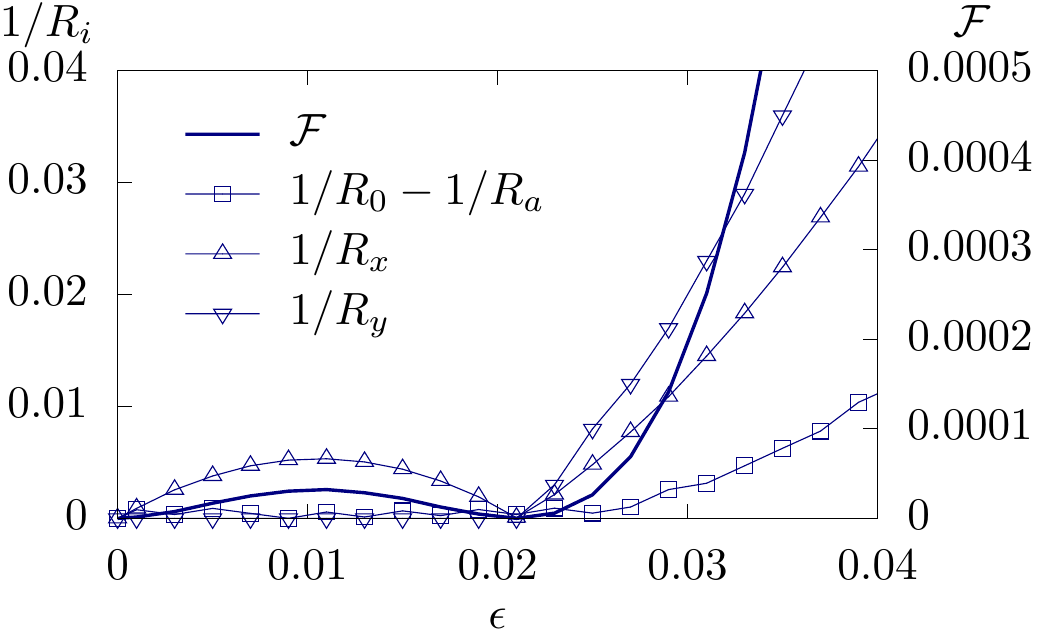}\\
	\caption{The elementary structure with two active (red, or dark) and four passive (yellow, or light) fibres. At $\epsilon<\epsilon_c$, a planar shape (a) is possible but has a higher energy than a non-planar state (b). At $\epsilon>\epsilon_c$, a planar shape (c) with all fibres bent to some degree is optimal. (d): The inverse optimal curvature radii of the parallel and perpendicular passive fibres, the deviation of the inverse curvature of the active fibres from its optimal value, and the residual energy of planar structures  as functions of the relative extension $\epsilon$. }
		\label{f:frame}
\end{figure} 

As a model example, we consider actuation by two active fibres along the $y$ axis in the minimal piece of textile with a single passive filament placed symmetrically parallel to the active ones and three passive filaments in the orthogonal $x$ direction connecting the ends and centres of the active filaments; all filaments have the same length $L$ before actuation (Fig.~\ref{f:frame}a--c). The integrity of this minimal structure is not supported by weaving, but we require the parallel and central perpendicular fibres to remain in contact at the central node, though still allowing for its sliding. Clearly, the planar shape cannot be retained at the conserved fibre lengths in either compressive or tensile actuation (at $\gamma_1=0, \,\gamma_2=\pi$ or \emph{vice versa}). An interesting case, allowing for an off-plane bifurcation is antisymmetric actuation with both active fibres oriented in the same direction ($\gamma_1=\gamma_2=0$). 

The general energy expression, scaled by $E r_a^4/\pi$ with $E=$ idem is dependent on the curvature radii of the active, parallel passive, and perpendicular framing and central fibres, respectively, $R_a$, $R_y$, $R_s$, and $R_c$, is 
\begin{equation}
\mathcal{F}= \frac {\rho^4}{8}\left(\frac{1}{R_y^2} 
+ \frac{2}{R_x^2} + \frac{1}{R_c^2}\right) + \chi^4 I_a (R_a),
 \label{engen}
\end{equation}
where $\rho=r_p/r_a$ and the coefficient $\chi \approx 1- \epsilon$ accounts for the change of $r_a$ upon actuation accounting for volume conservation. 

It is convenient to define arc angles $\phi_i=L_i/R_i$ with the indices $i \to (a,y,x,c)$ and the lengths $L_a=(1+\epsilon/2)L$ and all other $L_i=L$. We will also need the chord half-lengths (half-distance between the ends) $d_i=R_i\sin \phi_i/2$ and heights $h_i =R_i(1-\cos \phi_i/2)$. The character of deformations depends on the sign of the difference $\delta =d_a - L/2$. The critical elongation corresponding to $\delta =0$ is $\epsilon_c=4 R \arcsin [L/(2R)]/L - 2$, which amounts, for example, to $\epsilon_c \approx 0.021$ at $L=2, \, r_a=0.1$ with the optimal curvature radius under these conditions  $R_a=R_0 \approx 0.849 \,r_a/\epsilon$. 

At $\epsilon<\epsilon_c, \, \delta >0$, elongation of active fibres upon actuation overcompensates the decrease of $d_a$ caused by bending. A flat configuration is possible under these conditions when the framing perpendicular fibres bend inward with the curvature radius $R_x$ satisfying $d_a-d_y = h_x$. This deformation reduces the half-distance between the active fibres to $d_x <L/2$, and therefore the central perpendicular fibre must also bend with the same curvature radius $R_x$ to preserve its length, while the parallel passive fibre remains straight. The bending should be in-plane (in either direction) to preserve the contact with the rectilinear parallel passive fibre at the central node, and lifting the central node off-plane would require extending the framing fibres.The energy minimum is determined by a trade-off between energies of perpendicular passive and the active fibres, leading to a slight deviation of the latter's curvature radius from the optimal value $R_0$ (Fig.~\ref {f:frame}a).

The energy can be, however, further reduced by rotating the active fibres off-plane around their midpoint in opposite directions by some angle $\psi$. This reduces the projection of the base of the active fibre on the original plane to $\widehat{d}_a=d_a\cos \psi$, so that the framing passive fibres would become rectilinear at $\widehat{d}_a=d_y$. This leads to the absolute energy minimum $\mathcal{F}=\frac 18 \rho^4/R_c^2$ at $R_a=R_0$, $\cos \psi=1/d_a$, with $R_c$ satisfying $d_c=L \cos \psi$ and all passive fibres except the perpendicular central one remaining straight  (Fig.~\ref {f:frame}b). 

When $\epsilon>\epsilon_c$, $\delta <0$, the equilibrium structure remains planar. The parallel passive fibre may bend in-plane and the active fibres reduce their curvature to lower the difference $d_y-d_a$. In one limit, the parallel passive fibre remains straight, and the perpendicular fibres bend with the same curvature radius $R_x$ as at $\delta >0$, with the only difference that the framing fibres bend now outward. The energy of this configuration, scaled by $E =$ idem and $r_p^4$, is $\frac 34 R_x^{-2}$. In the opposite limiting case $R_y=R_a$, and the perpendicular fibres remain straight. The optimum lies between the above limiting configurations, so that all fibres are bent, as seen in Fig.~\ref {f:frame}c.

\subsection{Antisymmetric actuation\label{S3a}}

\begin{figure}[t]
\begin{center} 
 \begin{tabular}{cc}
		(a)&(b)\\
 \includegraphics[width=0.24\textwidth]{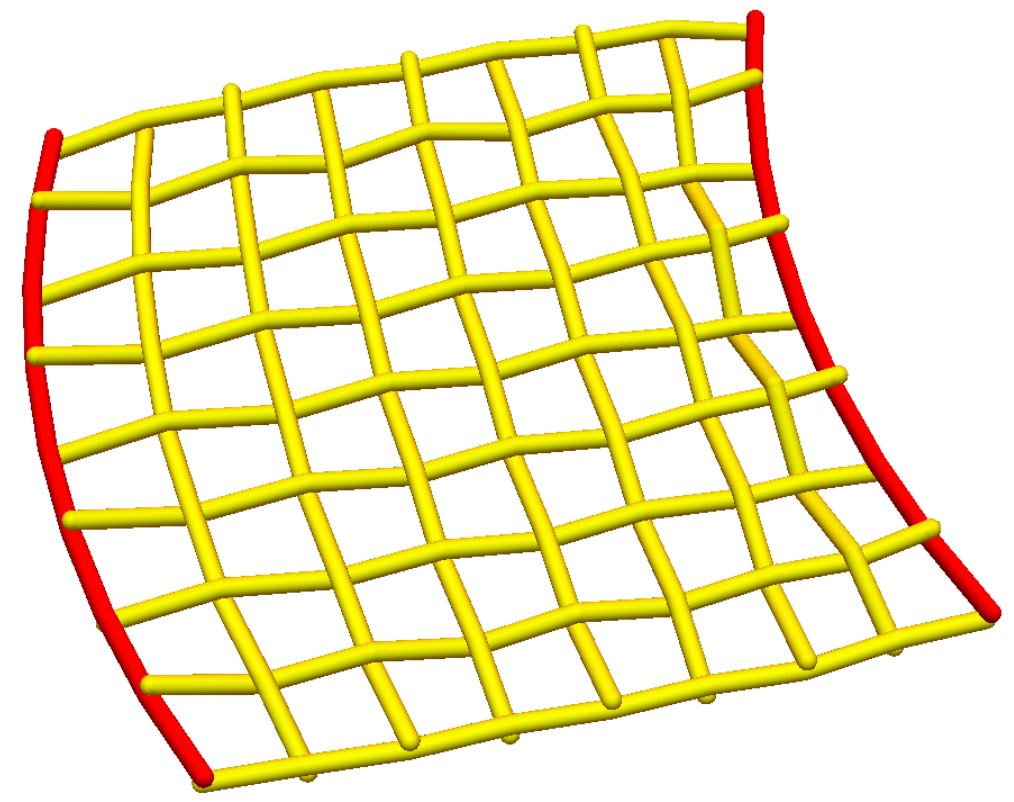} &
\includegraphics[width=0.24\textwidth]{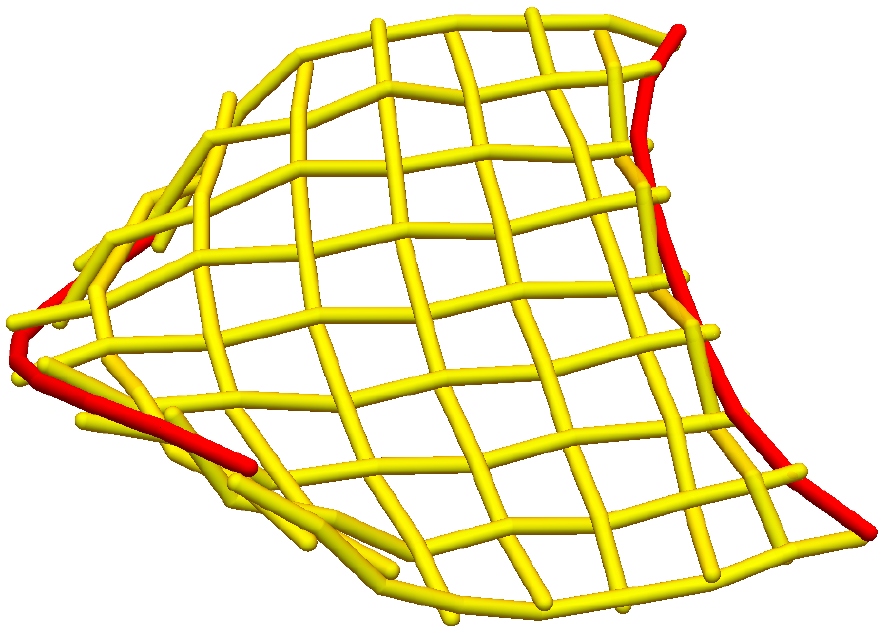}
\end{tabular}\\
 (c)\\ \includegraphics[width=0.4\textwidth]{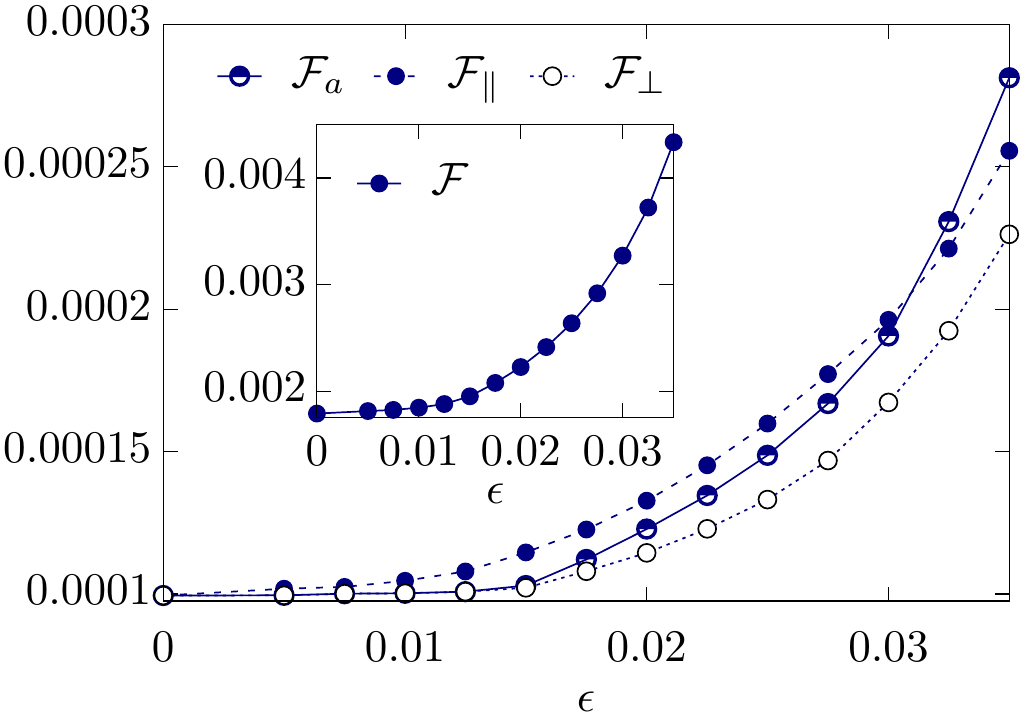}
\caption{Antisymmetric actuation. Examples of a planar (a) and non-planar (b) shapes for a square textile piece made of nine passive fibres in each direction with the unit mesh size, flanked by 2 active fibres; ${r}_a = {r}_p = 0.1$. (c): The dependence of the overall energy (inset) and average energy of the active, parallel and perpendicular passive fibres on the extension parameter $\epsilon$ in the optimal configuration at ${r}_a = {r}_p = 0.1$. }
\label{f:antisym}
\end{center} \end{figure} 
The competition between bending energies of active and passive fibres emerges as well in woven structures consisting of a larger number of filaments. 
We take as the initial configuration a flat square with two identically oriented active yarns embedded at the opposite edges. Unlike the above simple example, we allow in this computation for different curvatures at each fibre segment between nodes and take also into account microcurvatures preventing a close approach of fibres. 

Upon actuation, the network becomes irregular, as the distances between parallel passive fibres come closer to the active filament on the one side and larger on the other side. This leads to a difference in curvatures of active fibres, and as a result the distances between their centres become shorter than  the distances between their edges. Rotation of the active fibres off-plane around the central perpendicular fibre reduces the overall energy. Typical planar and non-planar actuated structures are shown in Fig.~\ref{f:antisym}a,b, and the energy as a function of $\epsilon$ is plotted in Fig.~\ref{f:antisym}c.

\subsection{Compressive/extensional actuation}\label{S3b}

\begin{figure}[t]
\begin{center}
\begin{tabular}{cc} (a) & (b)\\
			 \includegraphics[width=0.22\textwidth]{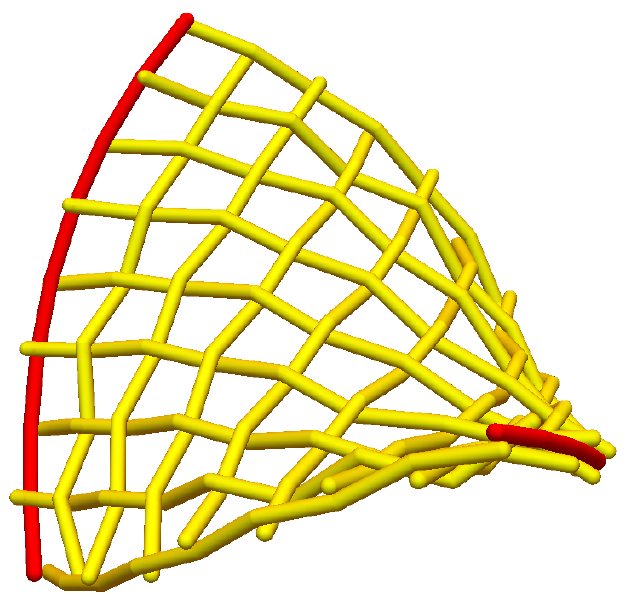}&
			  \includegraphics[width=0.22\textwidth]{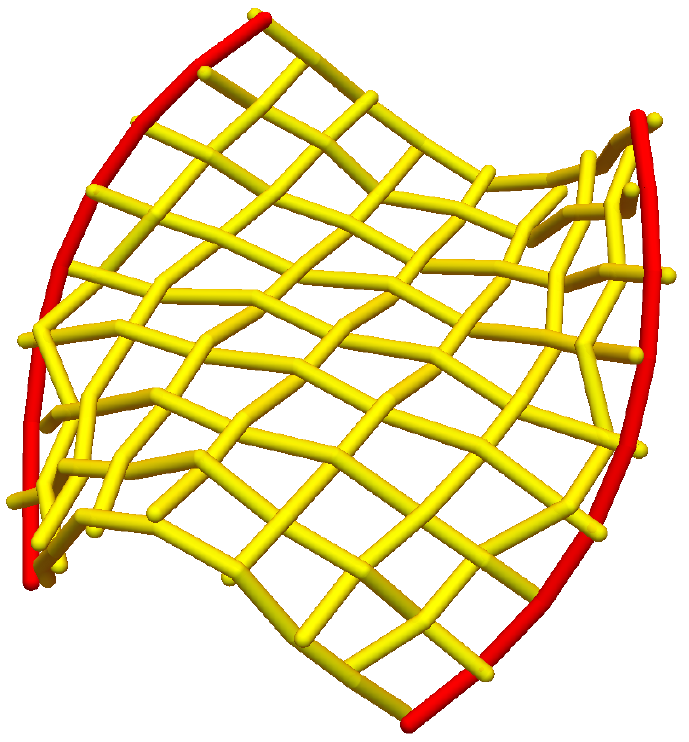}\\
			   (c) & (d)\\    \end{tabular}
			  \includegraphics[width=0.48\textwidth]{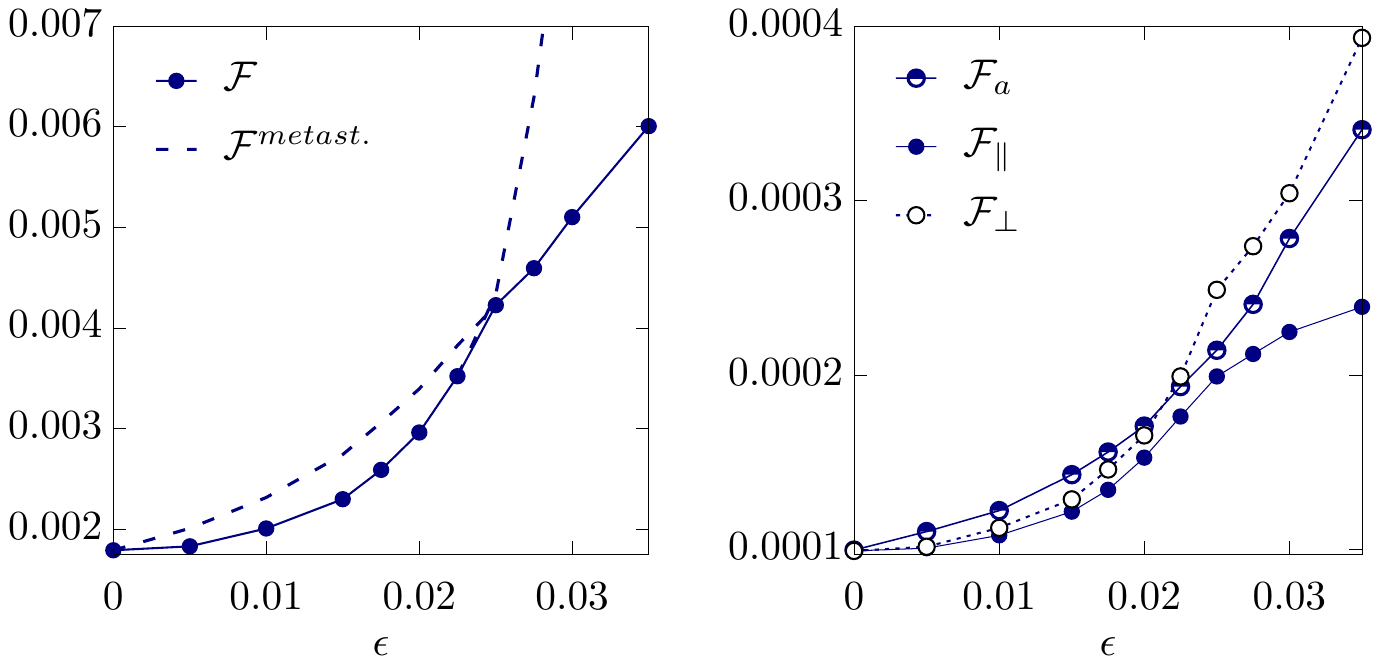}
	\caption{Actuation at $\gamma_1=\pi,\gamma_2=0$.  (a): A saddle shape at $\epsilon=1.0225$; (b): a ``monkey saddle" at $\epsilon=1.035$.  (c): The overall energy of both configurations as a function of $\epsilon$. The absolutely stable state is shown by the solid line and the metastable state, by a dotted line. (d): The dependence or the average energy of the active, parallel, and perpendicular fibres on$\epsilon$. All shapes and plots are computed at ${r}_a = {r}_p = 0.1$. }
	\label{f:Shape_180}
\end{center} 
\end{figure}

When both curvature radii are oriented in-plane inward ($\gamma_1=0,\gamma_2=\pi$),  the fabric experiences compression near the central line and stretching closer to the edges. Upon actuation, the  shape may deform in different ways. The two states have different energies, and the transition from absolute stability to metastability takes place when the extension parameter $\epsilon$ changes.
The bistability becomes even more pronounced if the active yarns are oriented outwards ($\gamma_1=\pi,\gamma_2=0$), causing stretching in the centre and compression near the edges. We will consider the latter case in detail.
       	
Evidently, the flat state cannot be retained upon actuation while preserving the lengths of the passive filaments. The two possibilities are either rotating the active fibres off-plane leading to a saddle shape as in Fig.~\ref{f:Shape_180}a or bending the passive fibres off-plane, with the curvature increasing toward the edges, leading to a ``monkey saddle" as in Fig.~\ref{f:Shape_180}b. In the former case, which is optimal at small $\epsilon$,  the residual bending energy of active fibres is comparatively low, and counterrotating the active fibres allows the passive fibres to remain almost straight. In the latter configuration, prevailing at increased $\epsilon$, the active fibres do not rotate and their curvature is close to optimal but passive fibres bend much stronger. 

The change of the overall energy for both configurations with $\epsilon$ at ${r}_a = {r}_p = 0.1$ is shown in Fig.~\ref{f:Shape_180}c, and the average energy of the active, parallel, and perpendicular fibres is plotted  in Fig.~\ref{f:Shape_180}c. The optimal shape switches with growing $\epsilon$ from a saddle to a ``monkey saddle" at $\epsilon\approx0.0225$. Both structures are stable, but non-optimal (metastable) states cannot be reached starting from the planar initial configuration. . 
      	               
\section{Off-plane actuation \label{S4}}
 
\subsection{Actuation at $\gamma_1 = \gamma_2 = \pi/2$\label{S4a}}

\begin{figure}[t]
\begin{center} 
 \begin{tabular}{cc}
		(a)&(b)\\
			 \includegraphics[width=0.25\textwidth]{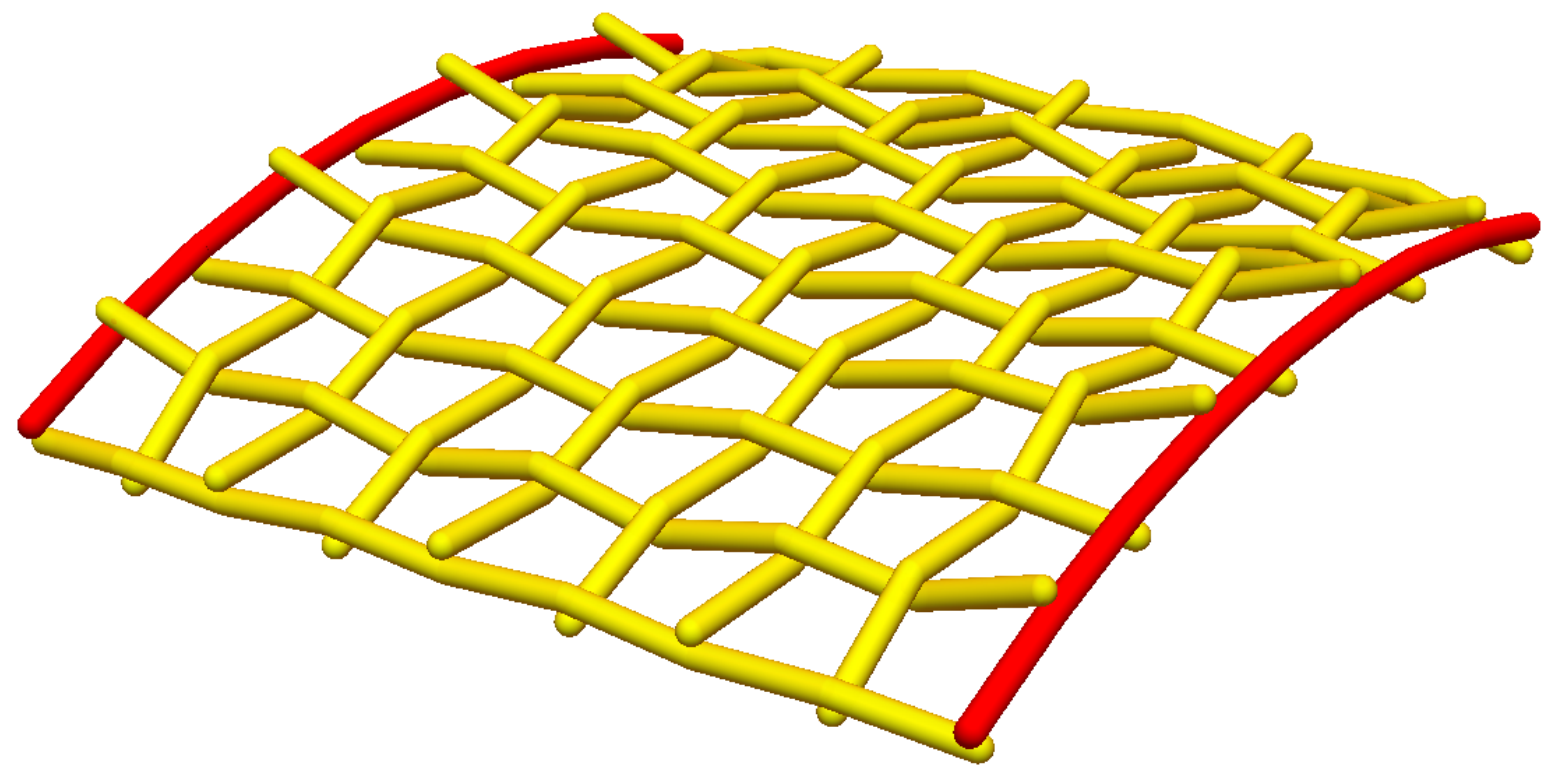}&
			 \includegraphics[width=0.22\textwidth]{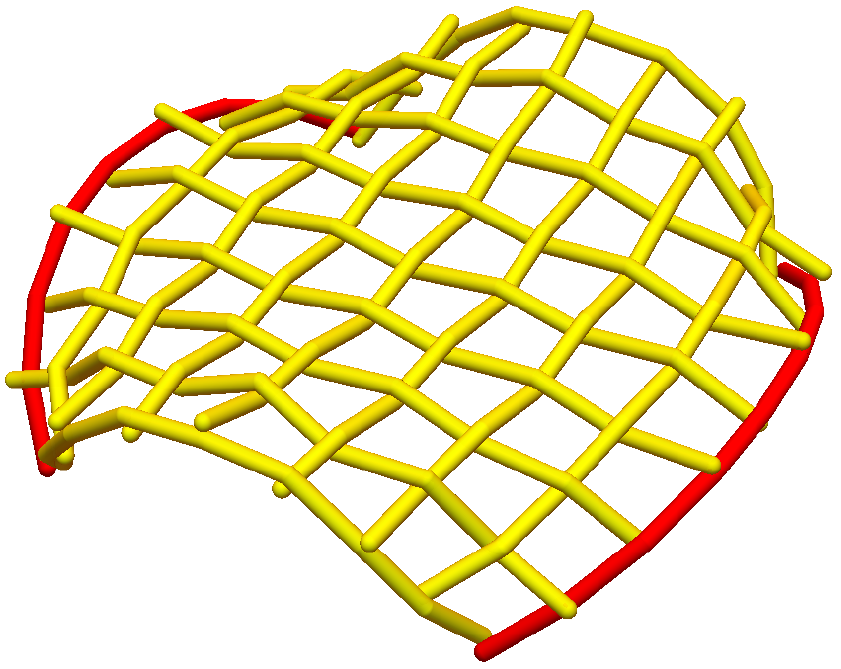}
			
\end{tabular}

\caption{Actuated states at $\gamma_{1}=\gamma_{2}=\pi/2$ and $\epsilon=0.06$ (a), $\epsilon=0.06$ (b). }
		\label{f:gamma}
\end{center} \end{figure}

If  the active fibres are oriented at $\gamma_1 = \gamma_2 = \pi/2$, so that they bend the textile in the normal direction, the parallel passive fibres resist strong bending. This causes the active fibres to turn their edges inward at higher elongations, thereby bending perpendicular passive fibres. Accordingly, the shape changes from that with a;most parallel active filaments and straight perpendicular passive fibres at small $\epsilon$ to that with non-parallel active filaments and bent perpendicular passive fibres at stronger extensions, as shown in Fig.~\ref{f:gamma}. The change is gradual in this case, with no sharp transitions.
     
\subsection{Venus flytrap}

\begin{figure}[h]
 \includegraphics[width=0.50\textwidth]{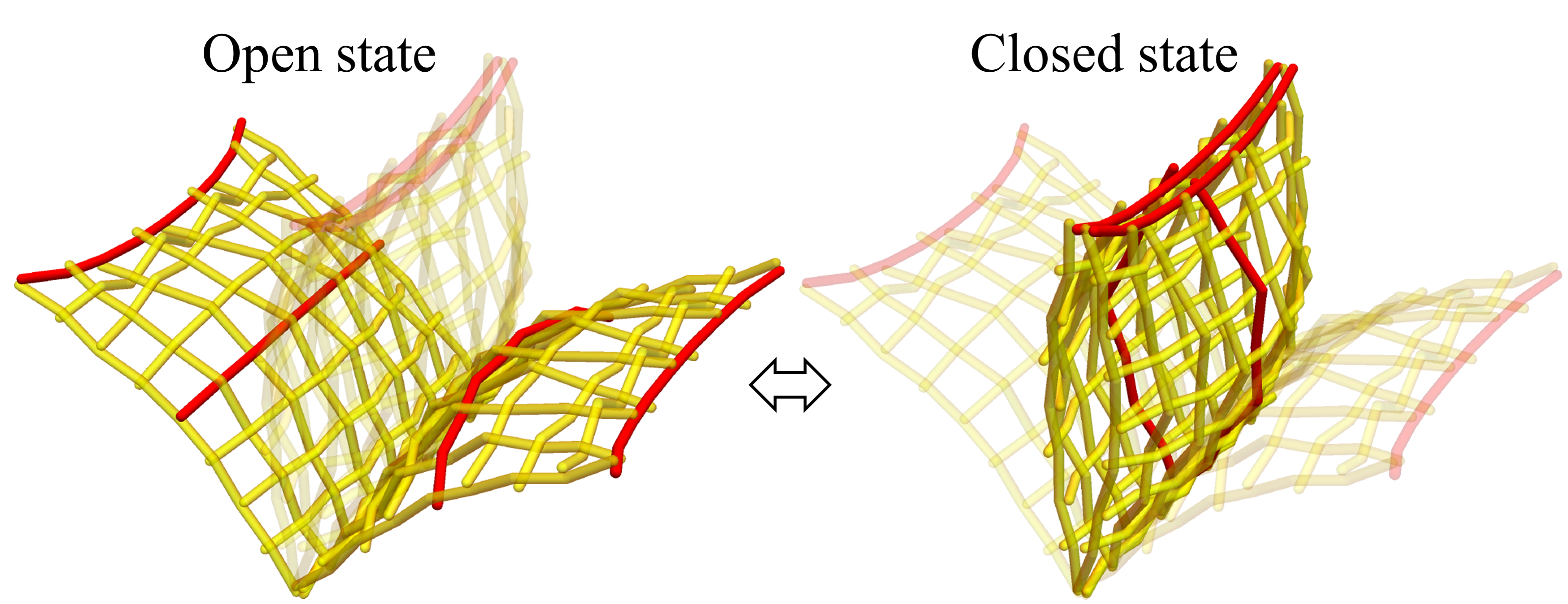}
	\caption{Fast closure of a flytrap by snap-through. Each half has one framing and two triggering active fibres in the centre. }
		\label{Flytrap}
\end{figure}

Sharp transitions are necessary for the snapping effect characteristic to Venus flytrap. Bifurcations to shapes of different character do occur under in-plane actuation (Sect~\ref{S3b}) but off-plane deformations cannot be in this case sufficiently large to create a closed shape. On the other hand, under off-plane actuation considered in Sect.~\ref{S4a} no sharp transitions are observed. A promising strategy is to combine both types of actuation. We start with two flat squares with a single Janus fibre each, attached along the side opposite to the active fibres and set at some relative angle. The flytrap is constructed by adding to the two framing active fibres (actuated in-plane and oriented inward) more active Janus fibres placed in the centres of both squares with the orientation of the curvature radius $\textbf{n}_1$ normal to the textile plane. Two added fibres are perpendicular and the other two are parallel to the framing active fibres, and both pairs are oriented in the opposite way normally to the plane of the fabric. Actuation of the framing active fibres reshapes each half to a symmetric state which can be either convex or concave. A fast switch between the alternative convex and concave shapes is implemented by actuating in turn the perpendicular fibres to close the flytrap and the parallel ones to open it, as it shown in Fig.~\ref{Flytrap}. This is similar to the action of the natural Venus flytrap with a leaf curved outward (convex) in the open state and inward (concave) in the closed state \cite{MahadevanFlytrap}. The angle between these artificial leaves can be fixed in the closed state, and remains constant during opening and closure of the flytrap. 

The shapes in Fig.~\ref{Flytrap} have the curvature radius up to the half of the fibre length, so that the angle between the open and closed states is about $\pi/2$. Increasing the extension coefficient $\epsilon$ will lead to a more curved shape but at the same time executing the snap-through will be more energetically expensive and is likely to be slower, requiring more time to the release elastic energy \cite{Vella}.

\section{Conclusion}

The above examples, though limited to simple structures (in all cases but the last one involving just a pair of active fibres) demonstrate a high versatility of active textiles. Extensive future studies, both theoretical and experimental, are needed to explore the full potential of this medium, that can be used to create dynamically evolving shapes and even allows for topological transformations of specially chosen configurations. We briefly noted in the Introduction many possibilities remaining outside the scope of our study. The versatility of textiles exceeds not only that of continuous sheets but also that of printed grids, as active fibres can be woven or embroidered in different ways, actuated independently and prepared with different internal structures incorporating varying composition and twist. 
\\
\\
\emph{Acknowledgement} This research is supported by Israel Science Foundation  (grant 669/14).


\end{document}